\documentclass[10pt]{wlscirep}

\usepackage[utf8]{inputenc}
\usepackage[T1]{fontenc}

\usepackage{pdfpages}
\usepackage{color,soul}
\usepackage{xcolor, ulem}
\usepackage{booktabs} 
\usepackage{tabularx} 
\usepackage{bm}
\usepackage{float}

\definecolor{highlightcolor}{HTML}{FFCC66}

\usepackage{lineno} 
\usepackage{ifthen}

\newif\ifreview

\reviewfalse

\ifreview
  \newcommand{\added}[1]{\sethlcolor{highlightcolor}\hl{#1}} 
  \newcommand{\deleted}[1]{\textcolor{red}{\sout{#1}}} 
  \linenumbers 
\else
  \newcommand{\added}[1]{#1} 
  \newcommand{\deleted}[1]{} 
\fi

\title{MIML: Multiplex Image Machine Learning for High Precision Cell Classification via Mechanical Traits within Microfluidic Systems}

\author[1]{Khayrul Islam}
\author[1]{Ratul Paul}
\author[1]{Shen Wang}
\author[2]{Yuwen Zhao}
\author[2]{Partho Adhikary}
\author[3]{Qiying Li}
\author[2]{Xiaochen Qin}
\author[4,2*]{Yaling Liu}

\affil[1]{Lehigh University, Mechanical Engineering and Mechanics, Bethlehem, 18015, USA}
\affil[2]{Lehigh University, Bioengineering, Bethlehem, 18015, USA}
\affil[3]{Lehigh University, Electrical and Computer Engineering, Bethlehem, 18015, USA}
\affil[4]{West China Hospital, Sichuan University, Precision Medicine Translational Research Center, Chengdu, China}

\affil[*]{yaling.liu@gmail.com}


\begin{abstract}
Label-free cell classification is advantageous for supplying pristine cells for further use or examination, yet existing techniques frequently fall short in terms of specificity and speed. In this study, we address these limitations through the development of a novel machine learning framework, Multiplex Image Machine Learning (MIML). This architecture uniquely combines label-free cell images with biomechanical property data, harnessing the vast, often underutilized biophysical information intrinsic to each cell. By integrating both types of data, our model offers a holistic understanding of cellular properties, utilizing cell biomechanical information typically discarded in traditional machine learning models. This approach has led to a remarkable 98.3\% accuracy in cell classification, a substantial improvement over models that rely solely on image data. MIML has been proven effective in classifying white blood cells and tumor cells, with potential for broader application due to its inherent flexibility and transfer learning capability. It is particularly effective for cells with similar morphology but distinct biomechanical properties. This innovative approach has significant implications across various fields, from advancing disease diagnostics to understanding cellular behavior.
\end{abstract}
\begin{document}

\flushbottom
\maketitle
%
%
\vspace{-7mm}

\thispagestyle{empty}

\section{Introduction}\label{sec:Introduction}

Identifying and sorting target cells from heterogeneous populations constitutes a crucial initial step in numerous biological, biotechnological, and medical applications \cite{Dainiak2007-uu, Shields2015-tp}. Following sorting, these cells may undergo detailed analysis, probing their proteomic, transcriptomic, or genetic identities and functions \cite{Baron2019-ue, Yousefi2023-mf, Banerjee2023-xd}. Alternatively, they can be utilized for regenerative medicine applications, such as transplantation into patients \cite{Stamm2003-uj, Bartsch2008-ol}. Cell sorting is traditionally performed based on molecular labels \cite{Miltenyi1990-xp, Bonner1972-nm, Shapiro2003-pz}. However, sorting methods leveraging intrinsic properties, such as cell size or deformability, have also been demonstrated \cite{Preira2013-dd, Wang2015-nb, Beech2012-qr}.

Nonetheless, it is imperative to acknowledge the limitations that emerge from the current approaches. While the high accuracy resulting from fluorescent methodologies is obtainable, it is not without its drawbacks \cite{Progatzky2013-ju, Cossarizza2021-tn, Basiji2007-xe, Ettinger2014-id, Li2019-mh}. The process of fluorescence labeling is both time-consuming and costly. Furthermore, fluorescent markers can interfere with cellular function and physiology, potentially altering their natural states and behaviors. Label-free bright field image-based cell detection techniques offer an alternative but often suffer from decreased classification accuracy, especially when distinguishing visually similar cells. \cite{Singla2023-vl, Bannur2023-mf, Siemenn2022-ey}

\added{Most existing machine learning-based approaches focus on either imaging data or extracted biophysical properties, often overlooking the complementary potential of integrating both modalities. For instance, some studies utilize scattering patterns, holographic imaging, or fluorescence data to classify cells based on specific features, while others rely on electrical or mechanical properties derived from microfluidic systems} \cite{Dannhauser2023-tr, Ahmad2022-qh, Dudaie2023-yk, Feng2023-dl, Nawaz2020-bh}. \added{Although a few studies have explored the intersection of label-free cell classification and machine learning} \cite{Chen2016-ij, Huang2021-bx, Liu2021-cm}, \added{to the best of our knowledge, none have fully integrated these two modalities into a unified framework. This approach is vital for developing advanced cell classification methods for visually indistinguishable cells, with the potential to transform our understanding of cellular dynamics. In this study, we address this critical gap by integrating imaging data and mechanical properties into a single machine learning pipeline, achieving state-of-the-art performance in classifying Human Colorectal Carcinoma (HCT116) tumor cells and White Blood Cells (WBCs), as detailed in Table 1.}

In our endeavor to refine cell classification methodologies, we developed a machine learning architecture called MIML. This architecture has been crafted to effectively tackle the technical complexities involved in classifying cells that are visually similar yet mechanically distinct, while also addressing the broader challenges associated with traditional diagnostic methods. The variability of observer interpretations and the scarcity of specialist pathologists are particularly acute in the field of cancer diagnostics and treatment monitoring. The availability of pathologists with specialized expertise remains limited even in well-resourced communities, giving rise to variability in diagnostic accuracy. This issue is intensified in underserved areas, where the deficiency of access to expert pathology services exacerbates diagnostic discrepancies, frequently resulting in prolonged or erroneous diagnostic outcomes that critically impair patient management and outcomes. MIML leverages both label-free bright field images and intrinsic cellular mechanical properties as primary input features, thereby incorporating cellular behaviors and states into the classification process. This approach minimizes the subjectivity associated with human observers and ensures that the results are reproducible and consistent. 

The effectiveness of the MIML model is demonstrated by differentiating tumor cells from WBCs. Detection of circulating tumor cells in blood samples is important for early cancer diagnosis and monitoring tumor progression \cite{Mikulova2011-ut, Wu2023-dt, Wu2022-yk}. Label-free detection without fluorescence labeling and antibodies is gaining popularity due to its applicability across various cancer types without the need for tedious labeling processes \cite{Zeune2020-yl, Wang2020-vj, Wang2024-sv, Dannhauser2020-ha, Maremonti2022-mf}. We used a model circulating tumor cell sample by mixing HCT116 cells, a well-characterized human colon cancer cell line, with WBCs to demonstrate that combining image features and mechanical features in an integrated machine learning framework leads to higher classification accuracy than using any single modality. By implementing MIML in this use case, we achieved a significant classification accuracy of 98.3\%, representing an improvement of approximately 8\% over pure image-based classification. With these insights and results in place, the core achievements of our study are summarized in the following research highlights:

\begin{enumerate} 

\item \textbf{Architecture}: We designed and introduced a new machine learning framework that uniquely combines image data and cell mechanical properties to predict tumor cells with high accuracy, circumventing the necessity for fluorescent labeling.

\item \textbf{Interpretability}: Through comprehensive feature analysis and activation layer visualization, we achieved an interpretable framework for tumor cell analysis, facilitating a deeper understanding and yielding higher accuracy in cell classification.
\item \textbf{Transfer Learning}: Our approach highlights the potential for transfer learning by successfully applying a combination of cell images and mechanical properties to classify visually indistinguishable cells and demonstrating its applicability to other cell types facing similar classification challenges.

\end{enumerate}

\begin{figure}[h]%
\centering
\includegraphics[width=0.8\textwidth]{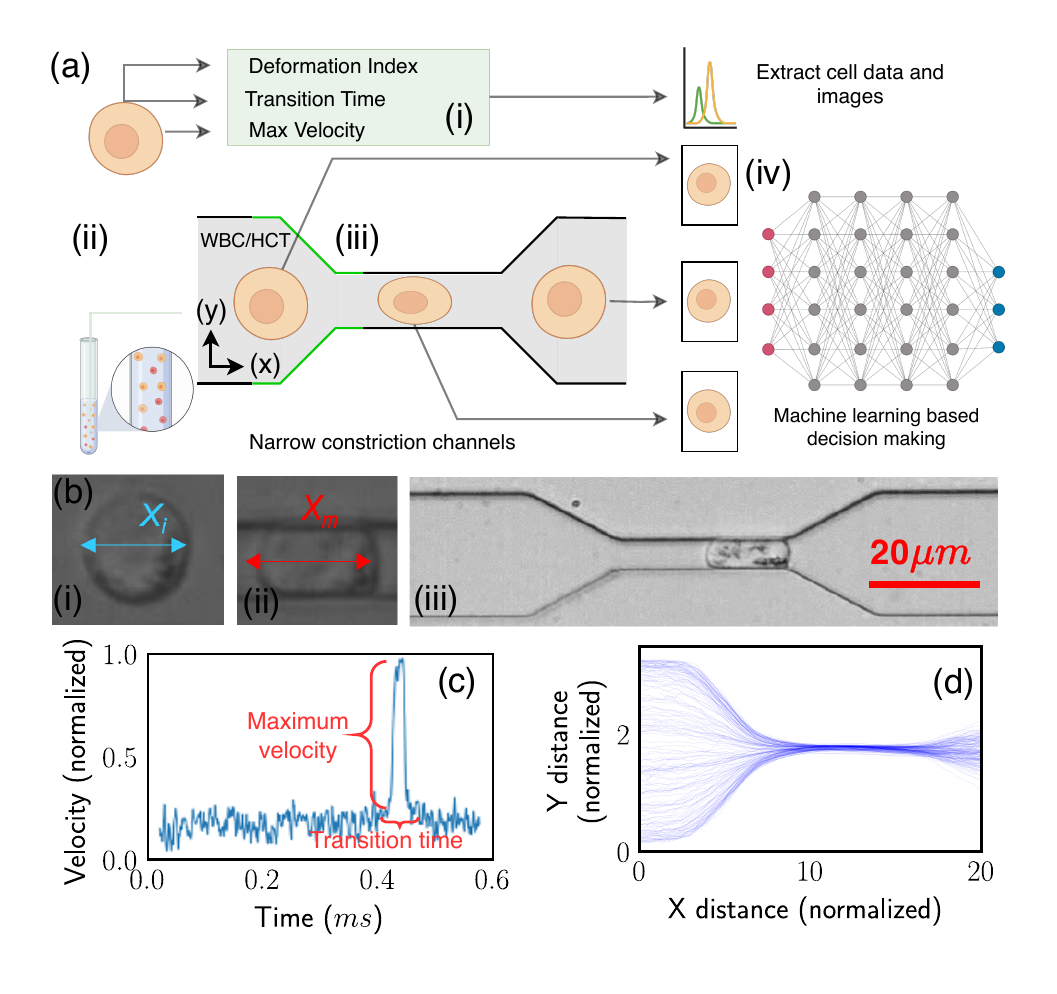}
\vspace{2mm}
\caption{MIML inferencing process and cell analysis. (a) Schematic of the cell data collection and subsequent classification, (i) biomechanical data collection, (ii) preparation of cell samples, (iii) cell transition through a narrow channel, (iv) MIML inferencing using cell morphology and feature, (b) experimental cell imagery, (i) cell length prior to compression, (ii) cell length while being compressed, (iii) snapshot of a cell positioned centrally within the narrow channel, (c) temporal velocity profile of cells, (d) normalized cell progression through the squeezing channel.
}\label{fig0}
\end{figure}

\section{Results and Discussion}\label{sec:Results}

The schematic representation of the standard procedure for cell detection via MIML is embodied in Figure \ref{fig0}(a). The subsequent sections are dedicated to a comprehensive examination of various classification models apt for numerical data. We also delve into traditional Convolutional Neural Network (CNN) models designed for image classification. Finally, we introduce our custom MIML model. The MIML model implements an innovative approach to predict cell type. It efficiently unifies images captured in real time as cells traverse through a narrow channel. The mechanical properties of these cells are evaluated through image processing techniques. This combination of visual and mechanical data allows for a more accurate and nuanced understanding of the cell type, thereby elevating the performance of cell detection and classification.

\begin{figure}[h]%
\centering
\includegraphics[width=0.9\textwidth]{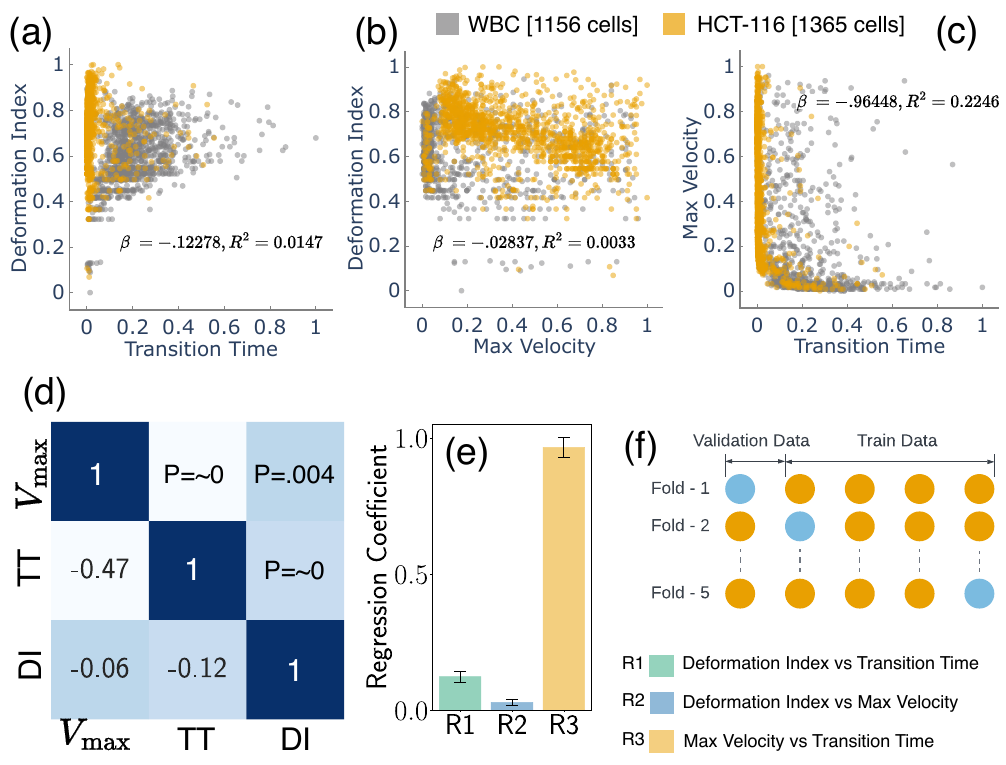}
\vspace{2mm}
\caption{Examination of cellular biomechanical properties. (a-c) Feature scatter plots that demonstrate the variability of inherent cellular attributes, (d) illustration of correlations between various features, as portrayed in a heatmap, accompanied by associated p-values, (e) regression coefficient between various features, (f) cross-validation utilized in this study, illustrating a scenario with four stacks used for training and one stack reserved for testing.}\label{fig1}
\end{figure}

\subsection{Composition of training and validation sets}\label{subsec:Results1}

In our study, we utilize two distinct forms of data for the purpose of training our models. The first, biomechanical feature data, is provided in a CSV format and serves as the training dataset for our classification models. The second, image data, is deployed in the training of our CNN model. To ensure robust model evaluation and to address concerns regarding the adaptability and precision of the algorithm with varied data sets, our data is partitioned into three sets: training, testing, and validation. Notably, the image data for the testing and validation sets is composed of recordings taken on different days, under diverse lighting conditions and camera settings, to simulate real-world variability and test the resilience of the algorithm against such fluctuations. The validation data is utilized to monitor model performance during training, while the testing data evaluates model efficacy post-training, helping to prevent overfitting, and ensuring the model generalizes well beyond training scenarios.\cite{Vabalas2019-ko, Lopez2014-pb}

To get the mechanical properties of the cells, we designed and fabricated a microfluidic device with a narrow channel smaller than the cell size. While the cells pass through such narrow channels, they will experience large deformation, leading to different translocation speeds and times. For each individual cell navigating through the narrow channel, we captured two images - one at the beginning and another at the termination of the squeezing process as shown in Figure \ref{fig0}(b)(iii). These images serve to train our CNN model, while an additional training data point per cell is employed for training our classification models. The total training data comprises 2521 entries, of which 1156 correspond to WBCs and 1365 to HCT-116 cells. For the purpose of training the CNN model, we possess a total of 5042 images, segregated into 2312 images of WBC and 2730 of HCT-116. The data corpus has been partitioned into training and testing sets with a 4:1 ratio. The larger portion (80\%) is allocated for training and validating purposes, while the remainder (20\%) is set aside for testing. This training set is further subdivided into five subsets for the implementation of cross-validation (detail on Results \ref{subsec:Results2}).

\subsection{Detection of cell classes by classification model}\label{subsec:Results2}

Image-derived metrics such as deformation, defined as the divergence from a perfectly circular shape, offer insight into the mechanical properties of the objects being measured \cite{Mietke2015-ex, Mokbel2017-en}. This deformation, coupled with maximum velocity and transition time, emerges as a crucial parameter for effective cell classification. We analyzed three distinct features of a cell navigating a narrow channel, namely, the cell's deformation Index $(DI)$, the Transition Time $(TT)$ taken by the cell to traverse the narrow channel, and the maximum velocity $v_{\text{max}}$ achieved by the cell within the channel's narrow region Figure \ref{fig0}(b)(iii). The DI of the cell is quantified using the following formula:

\begin{equation}
 DI = \frac{(a-b)}{(a+b)}
\label{eq1}
\end{equation}

\vspace{5mm}

In this equation, $a$ and $b$ represent the major and minor axes of the considered cell, respectively. The deformation index adopts a scale from 0 to 1, with 0 representing a flawless circle, thus indicating no deformation, and 1 corresponding to the utmost conceivable deformation, given an assumed minor axis of zero\cite{Rodrigues2015-wr}. As the deformation index values come pre-normalized, it is crucial to extend this normalization to the remaining pair of features - transition time and maximum velocity. By doing so, we amplify our model's generalization potential and encourage its adaptability across different contexts through transfer learning. Therefore, we ensure these features also conform to a 0-1 scale and proceed to scrutinize their correlation. 


Figure \ref{fig0} provides a detailed representation of the data acquisition methodology complemented by the microfluidic channel employed for cellular deformation. The channel includes a narrow constriction akin to a small tunnel, where the shortest gap from top to bottom measures 5.2 µm, and the front-to-back depth that the cells travel through is ~20 µm. This constriction size was specifically chosen to effectively challenge cell deformability, providing insight into their physical properties. \added{Tumor cells exhibit unique mechanical properties that distinguish them from normal cells, primarily due to cytoskeletal reorganization, which enhances their deformability and adaptability} \cite{Rother2014-wk, Rianna2020-bl, Li2017-pu}. For the cells under investigation, dimensions have been measured, revealing a mean diameter of 14.2 ± 4.4 µm for the HCT116 cell and a slightly smaller mean diameter of 13.5 ± 1.5 µm for WBC. The significance of these dimensions is that they directly influence how the cells interact with the microfluidic channel's constraints. The velocity, as delineated in Figure \ref{fig0}(c), is normalized in accordance with the flow velocity. Meanwhile, Figure \ref{fig0}(d) depicts the position of the cell's centroid, which is presented in values normalized relative to the cell's dimensions. This figure effectively captures the trajectory of the cell within the microfluidic channel. Figure \ref{fig1}(a-c) illustrated scatter plots capturing the relationship between each pair of feature sets under study. \added{A notable observation from Figure} \ref{fig1}(a) \added{is that HCT116 cells consistently exhibit shorter transition times and a greater deformation index ($\mu = 0.722$, $\sigma = 0.085$) as compared to WBCs ($\mu = 0.660$, $\sigma = 0.091$) with $p < 0.001$.} The underlying reason for this distinction lies in the inherent biophysical properties of these cell types. HCT116 cells are characterized by a lower stiffness \cite{Connolly2020-qe, Varol2022-lp}, which confers upon them a higher degree of flexibility. This malleability allows these cells to adapt their shape more readily in response to external pressures, consequently enabling a swifter transit through narrow spaces. This quality not only enhances their overall velocity but also results in a decreased transition time, albeit at the cost of experiencing greater deformation. The same reasoning can explain the lower maximum velocity of the WBC due to its reduced deformability. The relative stiffness of WBCs impairs their ability to modify their shape optimally to navigate the physical constraints of the narrow channel. This limitation subsequently slows down their transit time, as they lack the shape flexibility which is needed to maintain higher velocities.

\begin{figure}[h]%
\centering
\includegraphics[width=0.9\textwidth]{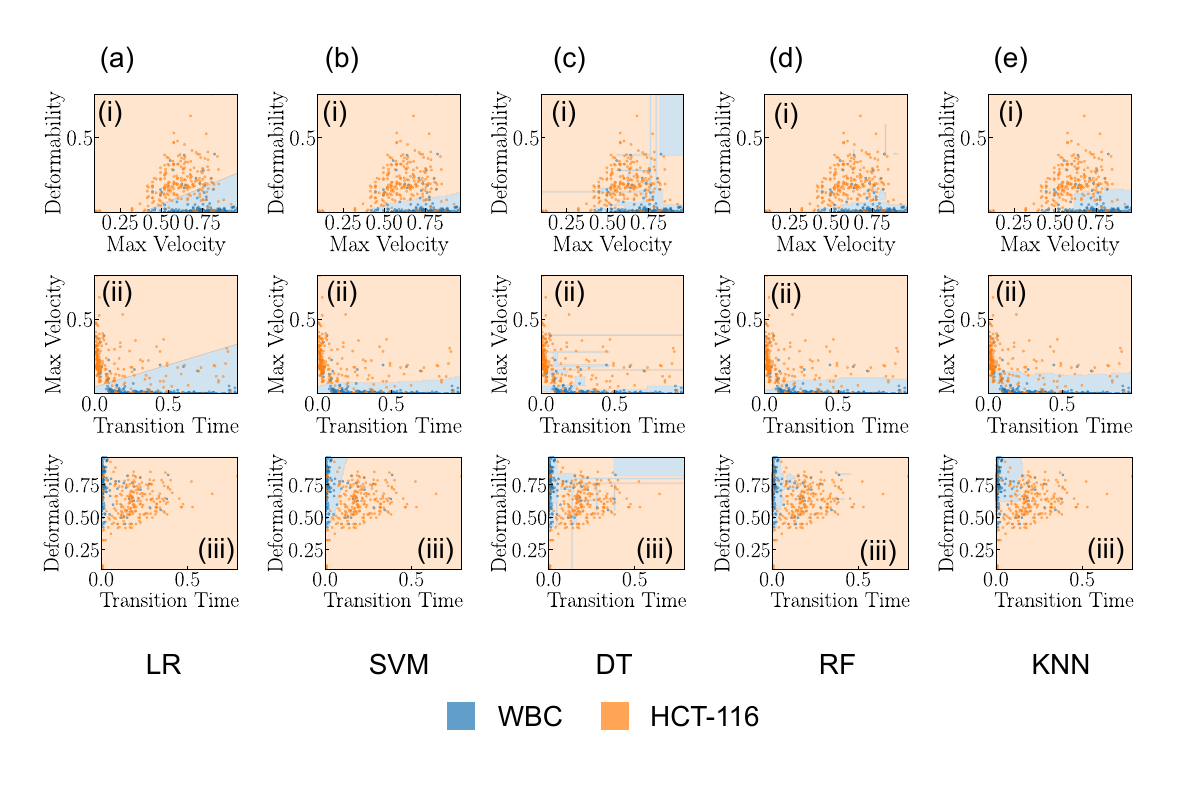}
\vspace{2mm}
\caption{Analysis of classification models. (a-e)(i-iii) Showcases the predictive capabilities of each classification model for varying feature combinations.}\label{fig2}
\end{figure}

Building upon these insights, Figure \ref{fig1}(d) presents a heatmap where the lower triangle shows the correlation matrix and the upper triangle indicates the associated p-values. Each p-value is a statistical metric that measures the strength of evidence for rejecting the null hypothesis in the context of evaluated features. Notably, all p-values displayed are less than 0.01, indicating strong evidence against the null hypothesis. This suggests the correlations observed are statistically significant and unlikely due to random chance. Even more compelling are two specific correlations (R1 and R3), with p-values $\sim 0$, implying an exceedingly strong level of statistical significance. To further investigate the training features, we plotted regression coefficients along with measured error bars. These coefficients provide a quantitative measure of the rate of change in one variable (dependent) due to a one-unit change in another variable (independent). Remarkably, the third relationship (R3), which is between maximum velocity and transient time, showcases a high regression coefficient of $\sim 0.96$. This high value suggests a robust association between these two variables, a finding that aligns intuitively with our understanding. As expected, the time needed for a cell to traverse through a narrow channel is inversely proportional to its velocity, an increase in velocity leads to a decrease in transit time, hence the strong relationship.

\begin{figure}[h]%
\centering
\includegraphics[width=0.9\textwidth]{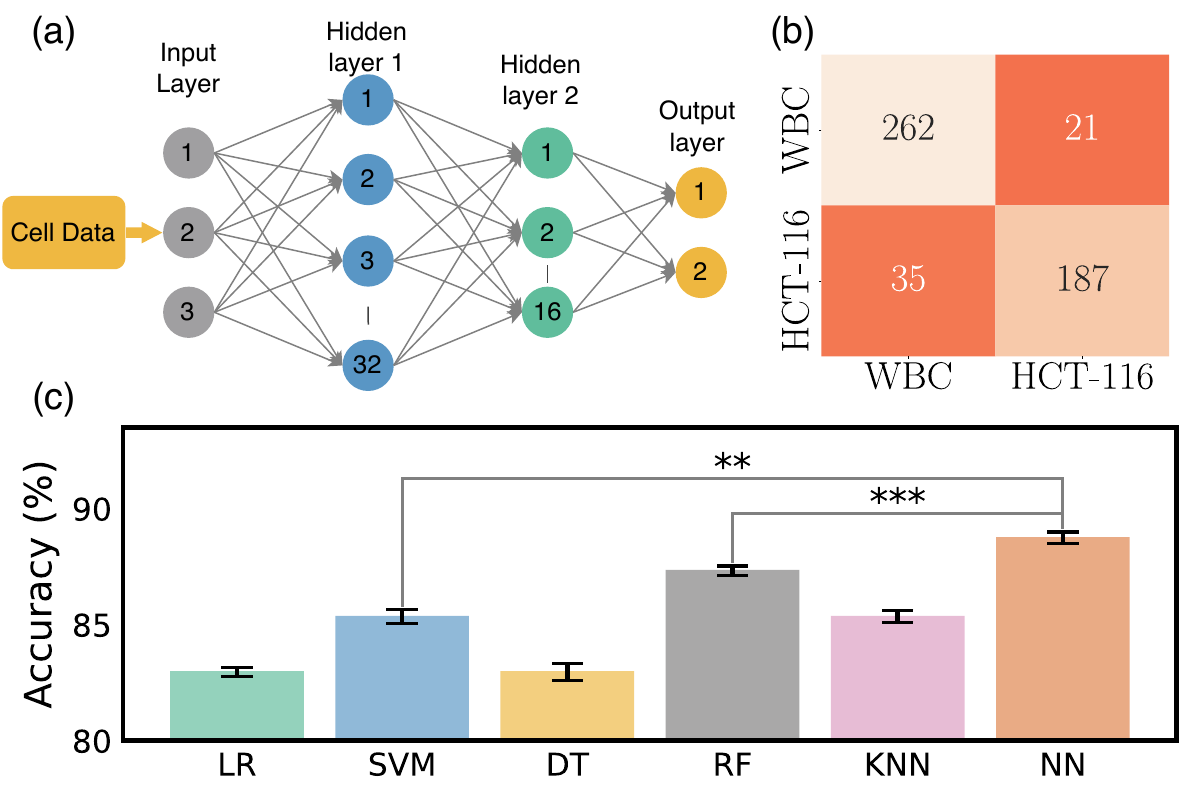}
\vspace{2mm}
\caption{Analysis of NN model. (a)Architectural design of NN dedicated to predicting cell classes, (b) confusion matrix, outlining the model performance against the testing dataset from the NN, (c) computed accuracy across all models. Statistical significance is indicated by ** ($p < 0.01$) and *** ($p < 0.001$)}\label{fig3}
\end{figure}

Figure \ref{fig1}(f) presents our cross-validation method. Four training stacks and one validation stack is used to prevent overfitting and accurately assess our model's performance and generalizability. After cross-validation, we evaluated the model on a separate testing dataset not used during training or validation. This strategy ensured the model's effectiveness on unfamiliar data, confirming the absence of biases from its training process\cite{Vabalas2019-ko, Xu2018-tj}. We then delved into the exploration of the predictive efficacy of several classification models, specifically Logistic Regression (LR), Support Vector Machine (SVM), Decision Tree (DT), Random Forest (RF), and K-Nearest Neighbors (KNN). Each of these models provides insights, as they are grounded in distinct computational approaches and theoretical underpinnings. LR, for instance, relies on statistical analysis to estimate probabilities, while the SVM employs geometric principles to maximize the margin between classes. DT and RF are built upon hierarchical structures that aim to split data into distinct subsets based on feature characteristics. Lastly, the KNN model classifies new instances based on their proximity to existing instances in the feature space\cite{Uddin2019-ul,Thanh_Noi2017-ql}. To get a nuanced understanding of the predictive power of these models, we employed each pair of features in our dataset to train these models. The results of these model predictions, excluding those from the Neural Network (NN), are visually represented in Figure \ref{fig2}(a-e)(i-iii). Through the systematic scrutiny of diverse models and feature combinations, we have obtained a comprehensive perspective of the predictive landscape inherent in our data.  We further extended our exploration to the prediction capabilities of NNs. Figure \ref{fig3}(a) delineates the architecture of the NN model that was employed in predicting cell types based on the features previously discussed. This architecture obtained through our ablation study (see supplementary material note 3 for detail) constitutes a simple yet effective network comprising two hidden layers containing 32 and 16 neurons, respectively. The first three input neurons are engaged in processing the input, while the final two neurons are tasked with generating the predicted class. To further quantify the performance of our model, we have also included a confusion matrix in Figure \ref{fig3}(b), which was derived from the testing dataset.

\begin{figure}[h]%
\centering
\includegraphics[width=0.9\textwidth]{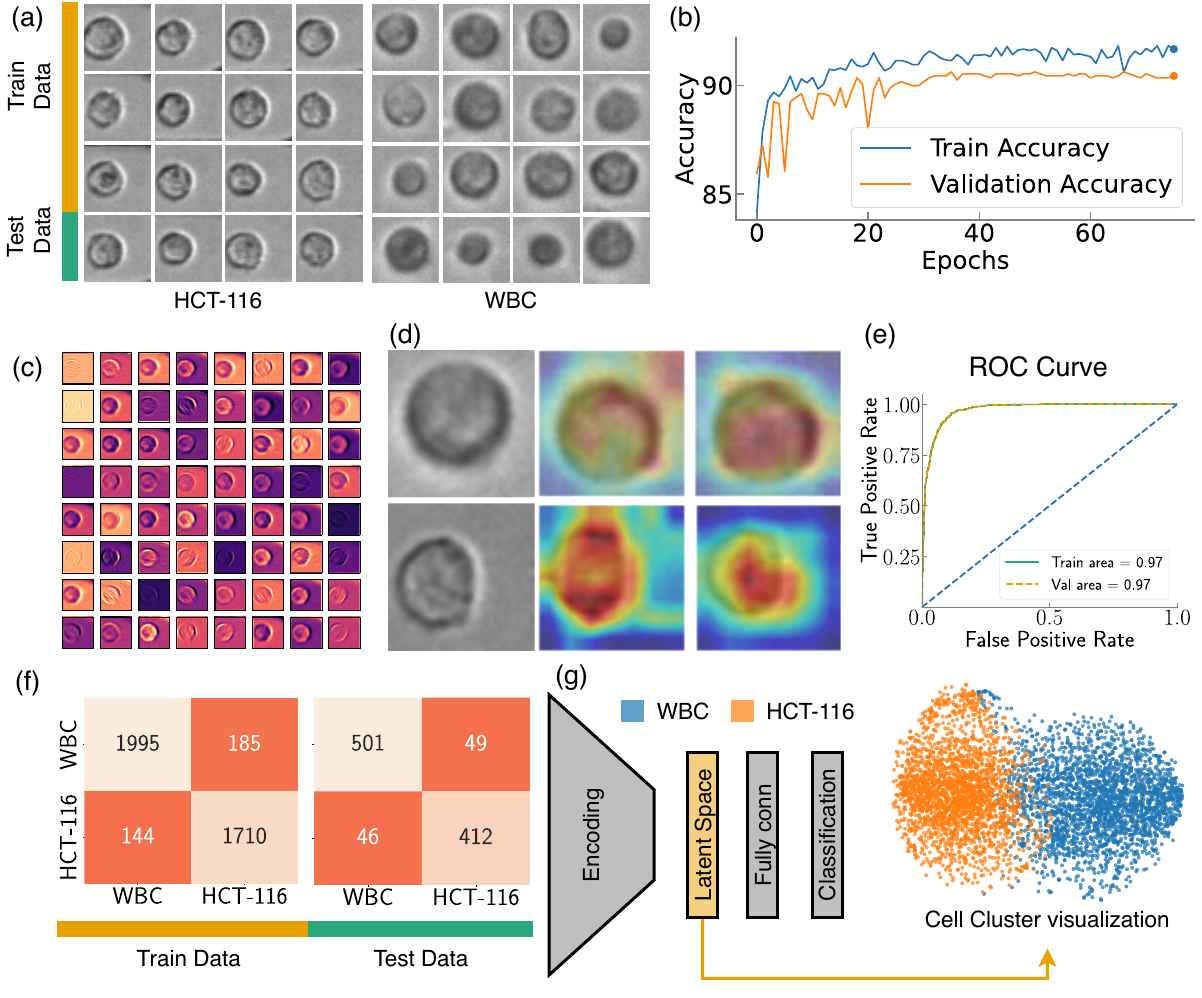}
\vspace{2mm}
\caption{Analysis of CNN. (a) Sample images from training and validation dataset, (b) training and validation accuracy while training, (c) feature map visualization for WBC, (d) Grad-CAM visualization for WBC and HCT116, (e) Receiver Operating Characteristic (ROC) curve, (f) confusion matrix, outlining the model performance against the training and validation dataset from the CNN, (g) illustration of our CNN architecture along with the TSNE plot generated from latent space.}\label{fig4}
\end{figure}

In line with this performance evaluation, Figure \ref{fig3}(c) showcases a comparative analysis of the accuracy of each evaluated model. Accuracy, defined in Eq. (\ref{eq2}), is the proportion of correct predictions relative to the total test datasets, serving as a key metric for evaluating model performance. From the visual representation, it is evident that the NN model exhibits the best performance compared to the other models we tested. The demonstrated capability of the system to accurately classify cell types based on specific feature sets reinforces confidence in its reliability and predictive accuracy. Given its performance, we decided to adopt the NN model for further exploration and classification of cells in our study.

\subsection{Detection of cell classes by Convolutional neural network}\label{subsec:Results3}

The CNN we used can be divided into two parts: a convolutional feature extraction part (called `Encoder') followed by the fully connected layers classifying the input based on the features (Figure \ref{fig4}(g)). The encoder is structured with an initial convolutional layer, succeeded by four progressive stages, each consisting of multiple blocks. These blocks are critical to the architecture, as they incorporate shortcut connections that perform identity mapping, with their outputs added to the outputs of the stacked layers. The shortcut connection is critical to solving the vanishing gradient problem, a common obstacle encountered during the training of deep NNs\cite{He2015-gi}. The initial layer is a 7x7 convolutional layer with a stride of 2, followed by batch normalization and a ReLU activation function, and finally max pooling. The four subsequent stages contain two blocks each, with the number of convolutional filters doubling at every stage, beginning from 64 filters in the first stage. Downsampling is performed by convolving with a stride of 2 in the first layer of the 2nd, 3rd, and 4th stages, excluding the shortcut connections, where 1x1 convolutions are applied to match the dimensions. Each of these convolutional layers is succeeded by batch normalization and a ReLU activation function. Following the final stage, there is a global average pooling layer and a fully connected layer that leads to the final classification output. The design of our CNN model, with its specialized blocks and skip connections, provides an efficient way to train deep networks by facilitating the propagation of gradients throughout the entire network. The accuracies obtained from the training and validation datasets have been visualized in Figure \ref{fig4}(b). As the figure illustrates, following 40 epochs of training, we observe a steady fluctuation in accuracy levels. This consistency in fluctuation suggests that the model reaches a relatively stable state of learning after the 40th epoch, indicating diminishing returns from further training. It's critical to highlight validation accuracy is almost the same as the training accuracy, implying that the model maintained a balance between learning from the training set and generalizing to the validation set. This observation is crucial in asserting the model's capacity to avoid overfitting providing a reliable and robust solution.

\begin{figure}[h]%
\centering
\includegraphics[width=0.9\textwidth]{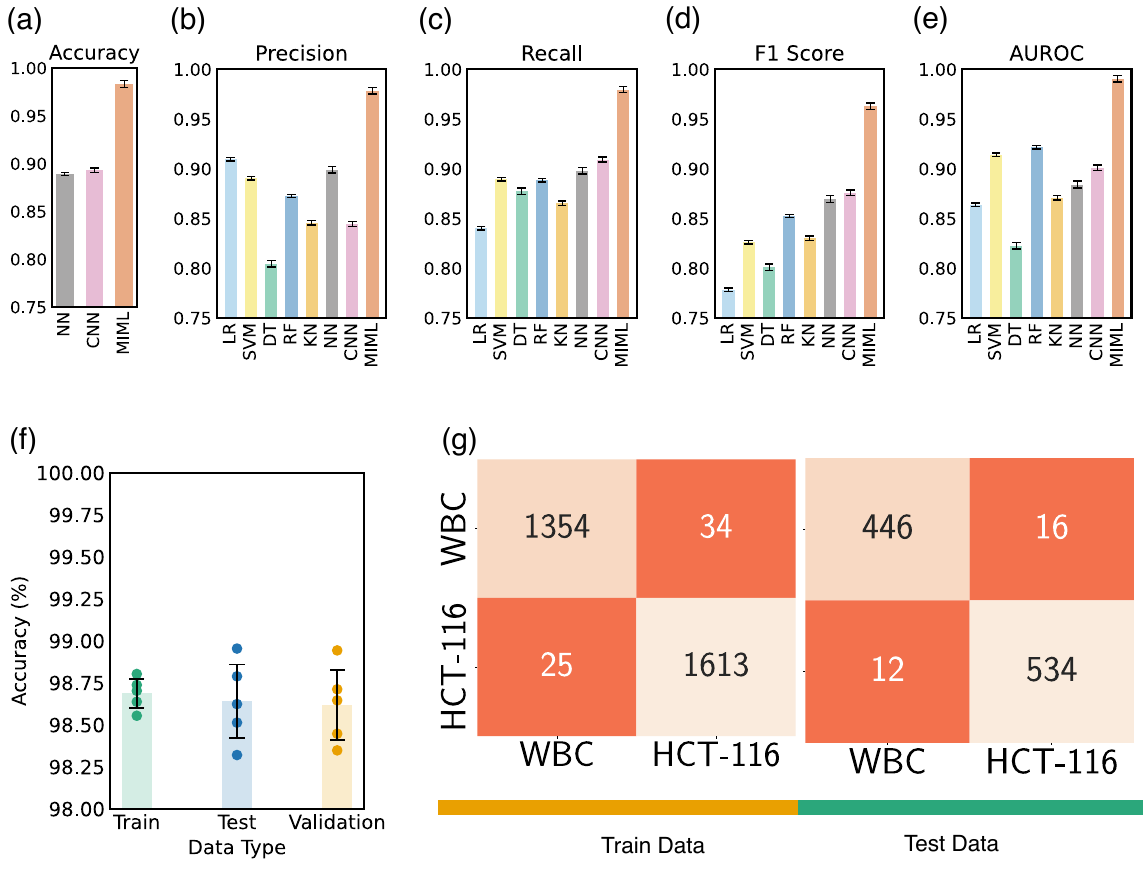}
\vspace{2mm}
\caption{Model Performance Evaluation. (a)-(e) Present the comparative assessment of Accuracy, Precision, Recall, F1 score, and AUC across various models; (f) showcases the training, testing, and validation accuracy specific to the MIML model, (g) displays the confusion matrix pertaining to both training and testing datasets within the MIML model.}\label{fig5}
\end{figure}

Figure \ref{fig4}(c) offers a detailed graphical representation of the first convolutional layer's feature map for WBC, incorporating all 64 filters of this layer. This feature map is an integral aspect of our study as it captures the distinctive features that the convolutional layer has learned to identify. Each of the 64 filters in this layer has learned to recognize different characteristics of WBC. For instance, some might specialize in detecting the contours, some may focus on textural information, while others might be zeroing in on more complex patterns. Visualizing these feature maps allows us to gain an understanding of the underlying mechanics of our model—what exactly it is picking up from the WBC images. By investigating these visualizations, we are essentially interpreting the model's learning process, which allows refining our model and augmenting its overall performance in the process\cite{Zeiler2013-av}. Equally important is the role of Gradient-weighted Class Activation Mapping (Grad-CAM) in model interpretability\cite{Selvaraju2016-wy}. In Figure \ref{fig4}(d), we illustrate its application, showcasing Grad-CAM for the penultimate and last convolutional layers, with a particular focus on HCT116 and WBC. The heat maps generated through the Grad-CAM are superimposed over the original images to provide a lucid understanding of the model's focus during its learning process. From the representation, it is evident that the penultimate convolutional layer is casting a broad net, capturing a substantial amount of background information, yet the primary emphasis remains on the cell structure. This layer acts as a broad filter, capturing both the cell and its surrounding context, which can be critical in many image recognition tasks. As the model progresses to the last layer, it significantly refines its focus. It zeroes in predominantly on the areas of the images that encapsulate the cell, displaying an acute understanding of the cell's morphology. This targeted approach underscores the layer's role in the identification of cell types based on their distinct morphological features. For the assessment of our model, we utilized the Receiver Operating Characteristic (ROC) curve, a crucial graphical tool for the evaluation of binary classification models, as illustrated in Figure \ref{fig4}(e). By graphically contrasting the true positive rate (TPR) against the false positive rate (FPR) at various decision thresholds, the ROC curve serves as a potent tool to measure the efficacy of our classification model\cite{Hanley1982-kq, Bradley1997-so}. In this instance, the ROC curves have been generated separately for both the training and testing datasets. Notably, the Area Under the Curve (AUC) manifests a remarkable consistency for both our training and validation datasets (Training area $\sim$ testing area). This consistency in the AUC values implies that our model exhibits no signs of overfitting - a common complication in machine learning where models tailor themselves too closely to the training data, compromising their capacity to generalize on unseen data. Instead, our model illustrates a balance between learning and generalizability, an unavoidable attribute in practical applications.

\begin{figure}[h]%
\centering
\includegraphics[width=0.9\textwidth]{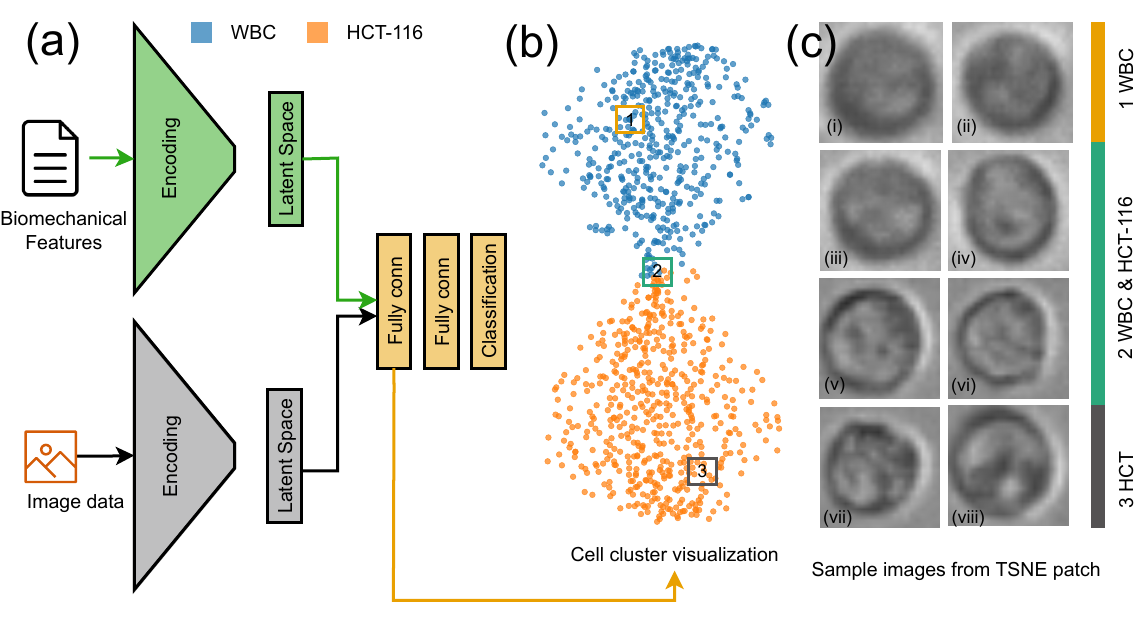}
\vspace{2mm}
\caption{Overview of MIML Cell Classification. (a) Schematic representation of the MIML architecture for cell class prediction, (b) t-SNE visualization derived from the latent space, (c)(i–viii) images sampled from marked patches in the t-SNE plot, showing actual dataset images of WBC(i-iv) and HCT116(v-viii)  cells to highlight visual differences at these latent space locations.}\label{fig6}
\end{figure}

Figure \ref{fig4}(f) offers a visualization of the predictions from our trained model in the form of a confusion matrix drawn separately for both the training and validation datasets. The results reveal a commendable level of accuracy for both datasets. The model's performance on the training data showcases an accuracy of $\sim 91.81\%$, demonstrating its effective learning from the given samples. Concurrently, the model has exhibited commendable performance on the validation dataset, attaining an accuracy of $\sim 90.6\%$. The validation accuracy is of particular importance as it indicates how well our model is likely to perform on unseen, real-world data. The close proximity of these accuracy values suggests a well-balanced model that has avoided overfitting, demonstrating robust learning from the training data while still maintaining the ability to generalize effectively to new data.
We extracted the features predicted by our model from the latent space and visualized them via a t-distributed Stochastic Neighbor Embedding (t-SNE) plot to unfold the high-dimensional data narrative (Figure \ref{fig4}g). The t-SNE is a robust machine learning algorithm, that excels in the visualization of high-dimensional data\cite{Maaten2008-lq}. It converts the similarities among data points into joint probabilities, endeavoring to minimize the Kullback-Leibler divergence between these joint probabilities in the low-dimensional embedding and the original high-dimensional data. This powerful technique provides a pathway to visualize the high-dimensional data captured in our CNN's latent space in a 2D format, offering an easily interpretable perspective. In our study, the latent space of our CNN model stored high-dimensional representations of the inputs, which encapsulated the abstract features that the model had learned. Transposing these representations into a t-SNE plot allowed us to take this complex, high-dimensional information and present it in a comprehensible, visually coherent format. Upon inspecting the t-SNE visualization, we observed a notable overlap between the clusters representing WBC and HCT116. This overlap suggests a visual similarity between these two cell types that the image-based CNN model has difficulty differentiating. This insight emphasizes the limitations of image-only models like CNNs in distinguishing intricate cellular characteristics and underscores the potential need for integrating other forms of data to improve cell differentiation performance.

\subsection{Multiplex Image machine learning for cell detection }\label{subsec:Results4}

To enhance the accuracy of cell classification, we developed a novel architectural model, aptly termed the Multiplex Image Machine Learning (MIML) Architecture. The MIML model processes both image data and mechanical property data for each cell passing through the microfluidic channel. For each training example, let \(\mathbf{x}\) be the input, represented as \((\mathbf{I}, \mathbf{m})\), where \(\mathbf{I}\) denotes the image of the incoming cell through the microfluidic channel, and \(\mathbf{m}\) represents the cell's mechanical property or feature vector. Here, the input image \(\mathbf{I} \in \mathbb{R}^{H \times W \times C}\) and the feature vector \(\mathbf{m} \in \mathbb{R}^d\), where \(H\), \(W\), and \(C\) are the height, width, and number of channels of the image respectively, and \(d\) is the dimensionality of the feature vector (in this case, \(d = 3\)). Each individual input \(\mathbf{x}\) is processed by the appropriate parts of the MIML model: the image \(\mathbf{I}\) is fed into a CNN, and the mechanical properties vector \(\mathbf{m}\) is fed into a NN. These components process their respective inputs separately and potentially combine their features at a later stage in the model. A batch of training data, denoted as \(\mathbf{X}\), consists of multiple training examples and can be represented as \(\mathbf{X} = \{(\mathbf{I}_1, \mathbf{m}_1), (\mathbf{I}_2, \mathbf{m}_2), \ldots, (\mathbf{I}_N, \mathbf{m}_N)\}\), where each pair \((\mathbf{I}_j, \mathbf{m}_j)\) corresponds to the image and mechanical properties of the \(j\)-th cell in the batch. The final classification output for a given example \(\mathbf{x}\) is denoted by \(\hat{\mathbf{y}}\) and is computed as:

\begin{equation}
\begin{split}
\centering
\hat{\mathbf{y}} = F_\text{MIML}(\mathbf{x})
\label{eq6}
\end{split}
\end{equation}

If \(\mathbf{z}_1\) is denoted as the resultant entity in the latent space emanating from the CNN, then it can be mathematically formalized as \(\mathbf{z}_1 = F_{\text{CNN}}(\mathbf{I})\). To optimally harness the high-dimensional abstractions synthesized by the CNN and to ensure their seamless integration with subsequent processing layers, the terminal fully connected (FC) layer is supplanted with an identity mapping. This substitution circumvents the final classification phase of the CNN, thereby facilitating the direct utilization of the latent embeddings, expressed as \(\mathbf{z}_1 = \mathbf{h}_1\), where \(\mathbf{h}_1\) signifies the high-dimensional feature vector distilled from the input image \(\mathbf{I}\) by the CNN. Concurrently, the mechanical property vector \(\mathbf{m}\) undergoes processing through a NN. Should \(\mathbf{z}_2\) be defined as the latent space output from the NN, it can be articulated as \(\mathbf{z}_2 = f_{\text{NN}}(\mathbf{m})\). Upon the derivation of \(\mathbf{z}_1\) and \(\mathbf{z}_2\) from their respective computational pathways, these vectors are concatenated to synthesize a cohesive feature representation, denoted as \(\mathbf{u} = F_{\text{Concat}}(\mathbf{z}_1, \mathbf{z}_2)\). In this context, \(\mathbf{u} \in \mathbb{R}^{n_{\text{cnn}} + n_{\text{mlp}}}\), indicating that the vector \(\mathbf{u}\) encapsulates the combined feature dimensions derived from both the CNN and NN pathways. The integrated vector \(\mathbf{u}\) is then propagated through a series of additional fully connected layers to achieve final classification. This involves a sequence of transformations: an initial linear transformation to an intermediate dimensional space, accompanied by batch normalization and ReLU activation; a subsequent linear transformation to a more compact feature space, again followed by batch normalization and ReLU activation; culminating in a final linear mapping that projects this refined feature vector onto the output class space, typically employing a softmax activation function for classification purposes. If \(h_2\) and \(h_3\) represent the dimensionality of the intermediate feature vectors \(\mathbf{h}_2\) and \(\mathbf{h}_3\), the class prediction $\hat{\mathbf{y}}$ can be defined as:

\begin{align}
\mathbf{h}_2 &= \text{ReLU}(\text{BatchNorm}(\text{Linear}(\mathbf{u}, h_2))) \\
\mathbf{h}_3 &= \text{ReLU}(\text{BatchNorm}(\text{Linear}(\mathbf{h}_2, h_3))) \\
\hat{\mathbf{y}} &= \text{Softmax}(\text{Linear}(\mathbf{h}_3, n_{\text{classes}}))
\label{eq7}
\end{align}

The strength of this integrated model lies in its ability to seamlessly handle and interpret both cell images and associated biomechanical features. As such, it encapsulates a broader perspective of cellular data, facilitating a more nuanced understanding and classification of the cells. Our empirical results underline the efficacy of the MIML model, yielding significantly higher accuracy levels in comparison to standalone implementations of the CNN or NN models. This improvement underscores the potential of leveraging multi-modal data – incorporating both image and biomechanical features – to substantially enhance the performance of cell classification tasks in machine learning applications.

As presented in Table \ref{tab1}, the enhanced performance of our MIML architecture becomes apparent through comparison with existing literature, including the study by Ozaki et al.\cite{Ozaki2019-ze}, which utilizes phase contrast imaging—a technique that, while detailed, is inherently complex and unsuitable for high-throughput applications like flow cytometry imaging. In contrast, our adoption of bright-field imaging simplifies the imaging process substantially and results in a notable improvement in both accuracy and F1 scores over the outcomes achieved by Ozaki et al\cite{Ozaki2019-ze}. and those reported by Piansaddhayanon\cite{Piansaddhayanon2023-qf}. These observations highlight the suitability of the MIML model for achieving efficient and precise cell classification. In our exploration of the efficacy of various machine learning models, including the MIML model, we investigate several key performance indicators. Specifically, we focused on metrics such as accuracy, precision, recall, the F1 score, and the area under the AUC as shown in Figure \ref{fig5}(a-e). Our evaluation framework was constructed around the aggregate count of true positives (TP), true negatives (TN), false positives (FP), and false negatives (FN) that were recorded during the model's predictions. These components form the foundation for our performance metrics, and their careful consideration is vital in the detailed dissection of our model's performance Eq. (\ref{eq2} - \ref{eq5}).

\begin{equation}
\begin{split}
\text{Accuracy} & = \frac{TP + TN}{TP + TN + FP + FN}
\label{eq2}
\end{split}
\end{equation}

\begin{equation}
\begin{split}
\text{Precision} = \frac{TP}{TP + FP}
\label{eq3}
\end{split}
\end{equation}

\begin{equation}
\begin{split}
\text{Recall} = \frac{TP}{TP + FN}
\label{eq4}
\end{split}
\end{equation}

\begin{equation}
\begin{split}
F1= \frac{{2\,\text{{Recall}} \times \text{{Precision}}}}{{\text{{Recall}} + \text{{Precision}}}} = \frac{2TP}{2TP + FP + FN}
\label{eq5}
\end{split}
\end{equation}

\vspace{7mm}

Our model showcases an accuracy improvement of $\sim$8\% compared to the pure image-based CNN model. As we delve deeper, we encounter precision, which allows us to zoom into the model’s positive predictions. Also known as the positive predictive value, precision quantifies the fraction of true positive predictions amidst all positive predictions made, a critical indicator when the implications of false positives are substantial. In this regard, our model surpasses CNN with a precision advantage of $\sim$8.4\%. Our exploration then pivots towards recall or sensitivity, another perspective-shifting metric that focuses on the actual positive cases, computing the proportion that the model correctly identifies. Its criticality surges when the repercussions of false negatives are high, ensuring that the model captures all relevant instances. Our model presents a recall improvement of $\sim$8.5\% over the alternative CNN model. Bridging precision and recall, we have the F1 score. This reconciling metric provides a balanced measure of a model’s performance by combining both precision and recall into a single entity. With its value oscillating between 0 (worst) and 1 (perfect precision and recall), the F1 score offers a comprehensive picture of the model’s performance. Herein, our model boasts an F1 score elevation of $\sim$8.7\% in relation to the CNN model. Finally, we engage with the AUC, a metric that transcends individual outcomes to evaluate overall model performance. It represents the probability that the model will rank a randomly chosen positive instance higher than a negative one. The true value of AUC shines as it evaluates both true positive and false positive rates, offering an all-encompassing performance view across all classification thresholds. Within this sphere, our model achieves an AUC enhancement of $\sim$9\%  compared to its counterpart. By analyzing these interconnected metrics, we elucidate the comprehensive performance profile of our models. Examining Figure 6(a-e), the superior performance of our MIML model is evident across all evaluation categories. This clear edge substantiates the importance of incorporating both image and cellular mechanical properties as inputs, demonstrating the efficacy of our approach. Notably, the significant leap in performance introduced by our MIML model suggests its effectiveness, underscoring a promising advancement in cell classification methodologies. 

\begin{table}[ht]
\centering
\caption{Comparison of the MIML Architecture's Performance with Published Studies in Cell Classification.}

\begin{tabularx}{0.8\textwidth}{@{} l X l l l @{}}
\toprule
Study                                               & Imaging Technique         & Model & Metric              & Score \\ 
\midrule
Ozaki et al.\cite{Ozaki2019-ze}.                    & Phase Contrast            & SVM               & Accuracy              & 94.2\%    \\
Piansaddhayanon et al. \cite{Piansaddhayanon2023-qf}& Bright-field              & CNN               & F1 Score              & 60.5 $\pm$ 0.4 \\
\midrule
\textbf{This Study}                                 & \textbf{Bright-field}     & \textbf{MIML}     & \textbf{Accuracy}     & \textbf{98.3\%}      \\
                            & \textbf{}             & \textbf{}                                     & \textbf{F1 Score}     & \textbf{96.3\%}      \\ 
\bottomrule
\label{tab1}
\end{tabularx}
\end{table}

\begin{figure}[H]%
\centering
\includegraphics[width=0.9\textwidth]{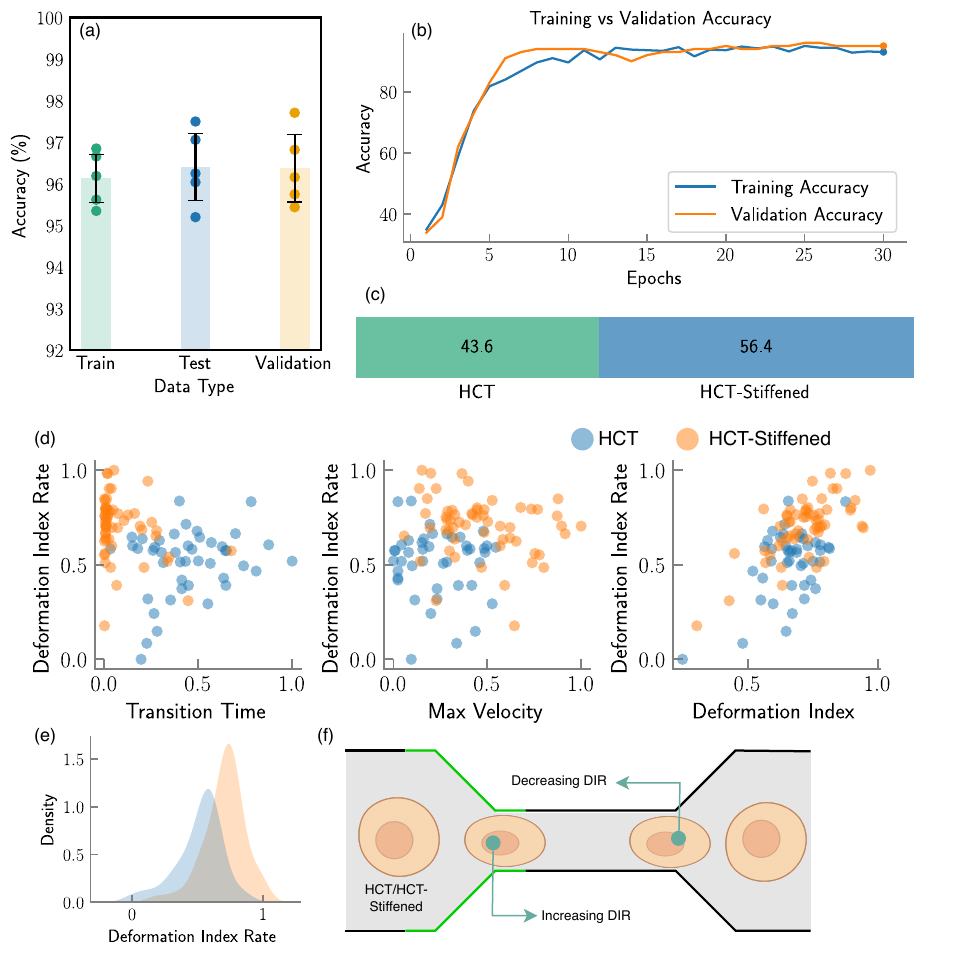}
\vspace{2mm}
\caption{Evaluation of MIML architecture on differentiating HCT vs vs. HCT-Stiffened Cell lines. (a) Comparative analysis of accuracy across training, testing, and validation phases, (b) evolution of training and validation accuracies with increasing epochs, (c) distribution of HCT vs vs. HCT-Stiffened cells within the dataset, represented as a percentage, (d) scatter plot comparing cell Deformation Index Rate (DIR) with other cell features, (e) Density plot of DIR for different cell types, (f) Schematic representation of a cell passing through a narrow channel, illustrating changes in DIR with varying cell stiffness.}\label{fig7}
\end{figure}

The model's applicability can be broadened to encompass additional use cases for more robust classification tasks by incorporating a greater number of cell features. For instance, as depicted in Figure \ref{fig7}(d), we introduced an additional feature, the cell Deformation Index Rate (DIR), to the existing feature set. The DIR is computed by averaging the rate of change of the deformation index over time, as the cell undergoes deformation and subsequently returns to its original shape through a narrow channel. The inclusion of this feature enhanced the detection accuracy by approximately 1.07\%, culminating in an overall accuracy of 98.3\%. Although this represents a modest improvement given the model's already high accuracy, incorporating this viscoelastic feature into the cell feature set holds the potential for substantial improvements in other use cases.

To provide a more nuanced evaluation of our MIML model, we have devised a composite visualization combining a bar chart Figure \ref{fig5}(f) and a scatter plot. The bar chart portrays the mean accuracies for the training, testing, and validation datasets derived from five-fold cross-validation, with each bar’s height representing the mean accuracy and the attached error bars denoting the variability in the results. Superimposed on this bar chart, we have a scatter plot that displays the individual accuracies from each of the five cross-validation trials. This layered presentation affords a more comprehensive overview of the model’s performance. The close agreement in the training accuracies across cross-validation trials reinforces the model’s reproducibility with the training dataset. Simultaneously, the proximity of the validation and testing accuracies to each other signifies the model’s robust generalizability, suggesting the absence of overfitting. 

We also investigated the diversity of the patterns presented in our data set. To do so, we randomly selected three patches within the latent space and visualized the images associated with them (Figure \ref{fig6}(i-viii)). Interestingly, despite an initial impression of high similarity among the images, the model was able to differentiate between them. On a superficial level, all three groups appeared quite alike, with subtle variations only perceptible upon meticulous examination. However, when processed through our model, these seemingly subtle differences were amplified, and the model distinctly categorized each group. Group 2, while almost indistinguishable from the other groups by eye, was identified by the model as possessing specific attributes that set it apart. Similarly, Group 1, and Group 3, despite their visible similarities to each other, were distinctly classified based on the model's analysis of the underlying patterns and structures in the data. This reveals the power and sensitivity of our model in distinguishing between seemingly identical data points. Even when human observers might struggle to discern any differences due to their apparent similarities, the model is capable of picking up on minute differences and categorizing the data accurately. This affirms the strength of our model in dealing with complex, high-dimensional data and underscores its potential utility in various fields where subtle variations in data could hold significant implications. The code necessary to replicate the MIML framework and adapt it to additional applications is available in our GitHub repository, which can be accessed at \url{https://github.com/BioNano-InterfaceLab/MIML}.

\subsection{Demonstrating MIML's Flexibility through Transfer Learning}\label{subsec:Results5}

This subsection delves into an experimental extension wherein MIML was tasked with distinguishing between two versions of the same cell line: a standard HCT116 (hereafter referred to as HCT) and a chemically stiffened variant of HCT116 (referred to as HCT-Stiffened). The stiffening of the latter was achieved through a specified chemical treatment, fundamentally altering its biomechanical properties without affecting its genetic makeup or morphological appearance. In this study, we utilized bright field microscopy images instead of fluorescence-activated cell sorting (FACS) due to the non-invasive nature and lower operational costs of bright field imaging, which allows for larger-scale data acquisition without the need for specialized dyes or reagents. Leveraging the pre-trained weights from the original MIML model—which was trained on a comprehensive dataset of $\sim 3,000$  cell images of WBC and HCT116 —we sought to explore the efficacy of the model in differentiating between HCT and HCT-Stiffened with a significantly smaller dataset. This approach not only underscores the model's inherent flexibility but also its capacity to generalize from limited information, a critical advantage in rapidly evolving research contexts. The training set for this transfer learning experiment consisted of  102 examples, which is approximately $\sim 30$ times smaller than the original dataset. Additionally, the model training stabilized within just 30 epochs, representing a reduction to nearly 50\% of the typical training duration. Despite the considerable reduction in data volume, the MIML model adeptly classified the HCT and HCT-Stiffened cells, achieving an accuracy of $\sim 96.5$. This level of accuracy, derived from a dataset substantially smaller than the one used to train the original model, exemplifies the potent transfer learning capability of the MIML architecture with the ability to adapt to new, albeit related, classification tasks by leveraging pre-existing knowledge, thereby reducing the need for extensive data re-collection and re-training.

\begin{figure}[h]%
\centering
\includegraphics[width=0.9\textwidth]{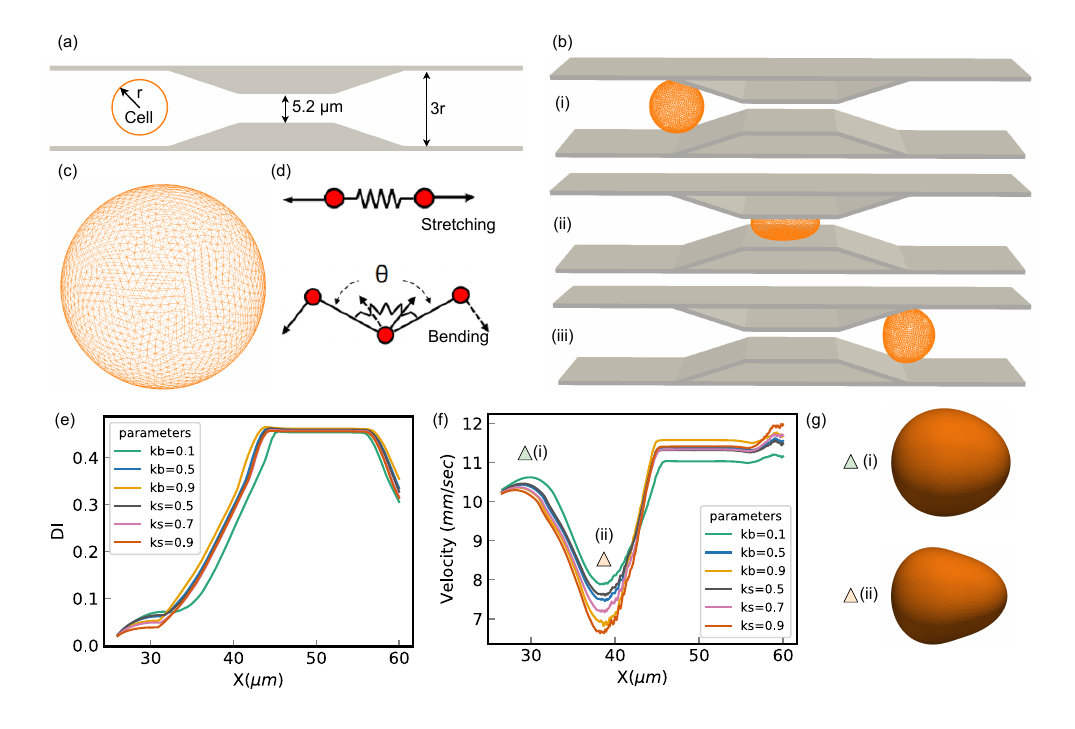}
\vspace{2mm}
\caption{ Overview of the Mesh-Based simulation approach. (a) Schematic representation of the computational domain, (b) temporal progression of cell deformation during transit through a narrow constriction, depicted at critical stages: (i) initial entry into the constriction, (ii) midpoint within the constriction, and (iii) exit from the constriction, (c) diagram of the spring-connected network modeling the cell membrane, (d) two-dimensional illustration of kinematics involved in stretching and bending within the spring-connected network, (e) plot of deformation index vs. x position, (f) Graph of velocity vs. x position, (g) snapshot of the cell at positions f(i) and f(ii)}\label{fig8}
\end{figure}

\subsection{Leveraging Simulation Data to Enhance MIML Training}\label{subsec:Results6}

While comprehensive datasets for various cell types are readily accessible online, acquiring corresponding mechanical data alongside cellular imaging remains a significant challenge. This limitation can hinder the effectiveness of the MIML framework for cell classification, which relies on both cell image and mechanical data. To bridge this gap and enhance the robustness and accuracy of the MIML framework, we have developed a simulation-driven approach that supplements the experimental data with synthetic but realistic datasets derived from computational models. These models simulate cellular mechanics, thereby generating dynamic cell mechanical data that complements the experimentally obtained cell mechanical data.

Our simulation framework utilizes a triangular mesh-based model to represent cellular structures (Figure \ref{fig8}(c,d)), capturing their biomechanical properties under various microenvironmental conditions. Employing the Lattice Boltzmann Method (LBM) as the flow solver, the framework robustly models the fluid dynamics that influence cell mechanics within microfluidic devices. The process begins with a warmup solution, establishing a steady-state flow field in the microchannel devoid of cells. This step is crucial to prevent abrupt and unphysical alterations in cell membrane dynamics or core mechanics due to unconverged fluid forces. Upon achieving a steady state, the main computational loop continues until the cell's minimum horizontal coordinates reach the channel's constriction. During this phase, the LBM solver updates the flow field, integrating forces and velocities with the cell model through a friction-based Immersed Boundary Method (IBM). This method is vital for realistically simulating the physical interactions between cellular structures and the surrounding fluid. The cells are discretized into networks of triangular patches, allowing for precise computations of local areal strain and corresponding cell mechanical property at each simulation timestep. This setup adeptly mimics the dynamic structural adaptations of cells under physical stresses, such as navigating through narrow channel constrictions (Figure \ref{fig8}(a,b)). Figure \ref{fig8}(e) illustrates the correlation between the DI and the cell center X-coordinate. This visualization underscores the variability in the DI as a function of the mechanical parameters $Ks$ and $Kb$, which are adjusted to emulate distinct cellular properties. The parameters $Ks$ and $Kb$ represent the stretching and bending coefficients, respectively, which dictate the magnitude of the force exerted by local triangular elements in response to deformations induced by stretching and bending forces. Figure \ref{fig8}(f) delineates the trajectory of cellular velocity until the point where the cell exits the constriction, facilitating the computation of the maximal cellular velocity and the transition duration. Furthermore, Figure \ref{fig8}(g) provides snapshots capturing the cell at the marked location in Figure \ref{fig8}(f), highlighting key velocity changes. The synthetic datasets generated from these simulations are crafted to supplement the experimental data, thereby amplifying the MIML framework's capacity for effective generalization from observed to unobserved data scenarios. This integrative approach not only broadens the dataset available for training the model but also deepens the understanding of cellular dynamics under diverse experimental and simulated conditions. To glean a comprehensive insight into the simulation methodologies and the intricate velocity profiles observed, readers are directed to our prior work in \cite{Nikfar2021-yd, Islam2023-vp}.

\section{Methods}\label{sec:Methods}

\subsection{Preparation of the microfluidic device}\label{subsec:Methods1}
The microfluidic channels were produced via the conventional UV lithography technique. Initially, channel designs were drafted using AutoCAD. Using the direct laser writing tool, DWL 66+ (sourced from the Quattrone Nanofabrication Facility at the Singh Center for Nanotechnology, University of Pennsylvania), chrome masks were created. These masks then facilitated the creation of the master pattern on an SU-8 2007 (MicroChem) layer on a silicon wafer, executed at the Center for Photonics and Nanoelectronics (CPN) at Lehigh University. The SU8-2007 was applied to the silicon wafer at a speed of 1000 rpm. Following a soft bake phase, the SU-8 underwent UV exposure using the Suss MA6/BA6. Post-development, the SU-8 designs underwent a hard bake at 150°C for 30 minutes. Sylgard 184 PDMS, combined with its curing agent at a 10:1 ratio, was poured onto the photoresist master. After a 2-hour degassing period for the PDMS, it was allowed to cure overnight in an oven. Finally, the inlets and outlets were created in the PDMS channel prior to its attachment to a large coverslip, secured with oxygen plasma treatment.

\subsection{Cell culture and data collection}\label{subsec:Methods3}
The Human Colorectal Cancer cell line (HCT116), was purchased from the American Type Culture Collection (ATCC), and cultured in Dulbecco's Modified Eagle's Medium (DMEM, Gibco). The DMEM was supplemented with 10\% Fetal Bovine Serum (FBS, Gibco), and 100 U/mL of Penicillin Streptomycin (R\&D system. To maintain the cells in an optimal state, we changed the culture medium every other day. To isolate single cells, we employed a 0.05\% Trypsin-EDTA (Bio-Techne Corporation. The HCT-116 cells were used in two formats (e.g., normal condition and stiffened condition). To fix the cells and make them stiffened, HCT-116 was treated with 4\% paraformaldehyde solution in PBS and subsequently incubated for 10-20 minutes. Then the cells were washed with PBS for experimentation. The Human Peripheral Blood Mononuclear cells (PBMCs) were obtained from the Human Immunology Core at Penn Medicine. These PBMCs were maintained in RPMI-1640 medium (ATCC), fortified with 10\% FBS and 100 U/mL of Penicillin-Streptomycin. All the bright field images were captured using a Nikon Eclipse TE2000S inverted microscope with an Ximea CCD Camera.  All the images were taken with the same setting for comparison.

\subsection{Feature extraction}\label{subsec:Methods4}
During our investigation, we implemented an experimental protocol that guided the cells through carefully constructed, narrow passages. Our ability to detect and scrutinize these cells was enhanced by a straightforward yet potent machine learning model - Yolov5. This model generated bounding boxes, a crucial tool that helped us concentrate on each individual cell, even when they were under the pressure of induced deformation. Within these defined areas, we initiated a detailed analysis of several parameters. As a first step, we computed the deformation index, a metric indicating the degree of cellular deformation. Concurrently, we methodically measured the time taken by each cell to travel the entire length of the channel, a metric defined as Transition Time Figure \ref{fig0}(c). We also evaluated the maximum velocity achieved by each cell during its passage through the channel. Upon calculating the aforementioned parameters, we segmented the cell image and stored it along with its measured property for subsequent training of our MIML model.

\section{Conclusion}\label{sec:Conclusion}

This study demonstrates that integrating label-free cell images with biomechanical properties using the MIML architecture significantly enhances cell classification accuracy, achieving a rate of 98.3\%. By capturing both morphological and mechanical traits, MIML provides a more comprehensive cell profile than models relying on a single data type. The successful classification of WBCs and HCT116 tumor cells underscores the model's effectiveness, particularly for cells that are visually similar but mechanically distinct. The implications of this approach are substantial. Improved classification accuracy can advance disease diagnostics, enable more precise monitoring of tumor progression, and facilitate the development of targeted therapies. Furthermore, integrating MIML with clinical workflows and leveraging clinical information—such as patient history, diagnostic imaging, and biomarker profiles—could provide complementary guidance for detecting cell-specific properties and enhance the interpretation of classification results. This synergy could not only improve diagnostic precision but also streamline the translation of MIML into practical medical applications. Additionally, the transfer learning capability of MIML suggests that it can be adapted to other cell types and applications, potentially transforming various biomedical fields. Future work should explore the application of MIML to a wider range of cell types and investigate its integration into clinical workflows. Addressing limitations such as the need for specialized equipment to measure biomechanical properties and exploring methods to incorporate readily available clinical data could further enhance the model's practicality. Overall, MIML represents a promising step toward more accurate and holistic cell classification methodologies.

\bibliography{main}

\includepdf[pages=-]{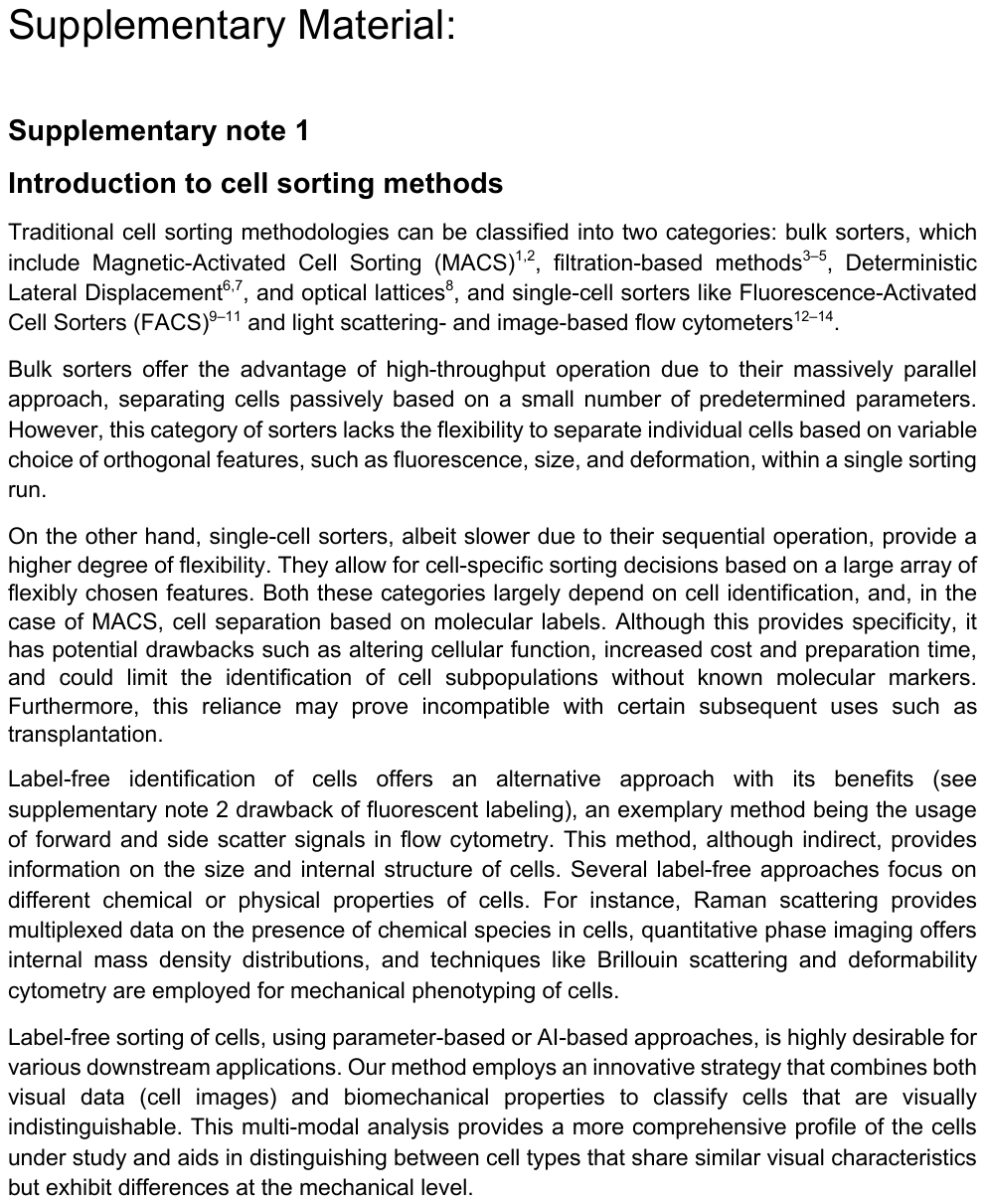}

\end{document}